\documentstyle[prl,aps,preprint]{revtex}

\begin{document}
\draft
\title{The addition spectrum of interacting electrons: Parametric 
dependence}

\author{Luca Bonci and Richard Berkovits} 

\address{
The Minerva Center for the Physics of Mesoscopics, Fractals and 
Neural Networks,\\ Department of Physics, Bar-Ilan University, 
Ramat-Gan 52900, Israel}

\date{\today}
\maketitle

\begin{abstract}
The addition spectrum of a disordered stadium is studied for up to 
120 electrons using the self consistent Hartree-Fock approximation 
for different values of the dimensionless conductance and in the 
presence and absence of a neutralizing background. In all cases a 
Gaussian distribution of the addition spectrum is reached for $r_s 
\leq 1$. An empirical scaling for the distribution width in the 
presence of a neutralizing background is tested and seems to describe 
rather well its dependence on the dimensionless conductance. 
\end{abstract}  

Recent measurements of the distribution of the addition spectrum of 
chaotic quantum dots~\cite {sba,shw,charlie} differ from the results 
of the orthodox constant
interaction (CI) model~\cite{Kastner,Meirav,Ashoori,McEuen} in 
several ways. The distribution, which is roughly Gaussian in all 
experiments, has no bimodal structure expected due to the 
spin~\cite{ber}, nor does it have a Wigner like distribution expected 
if somehow the spin degeneracy is lifted~\cite{Blanter,ba}. The width 
of the distribution varies between the different experiments. While 
in the first experiment~\cite{sba} the width is considerably larger 
than that expected from the CI model, the width in the latest 
experiment~\cite{charlie} is compatible with the CI model predictions.

These deviations from the CI model are usually attributed to the low 
electronic densities of the dots. For a typical dot the electronic 
density is of the order of $n_D = 2 - 3.5 \times 10^{11} \ {\rm 
cm^{-2}}$, corresponding to a ratio of the average inter-particle 
Coulomb interaction and the Fermi energy $r_s=1/\sqrt{\pi n_D} a_B 
\sim 1$ (where $a_B$ is the Bohr radius). At these values of $r_s$ 
correlations in the electronic densities which are not taken into 
account in the CI model appear~\cite{sba,ba}. 

Thus, the study of the addition spectra for chaotic dots must take 
into account the intricate interplay of chaos and interaction. 
Several numerical studies have tried to clarify the picture using 
exact diagonalization methods~\cite{sba,ber} and Hartree-Fock (HF) 
approximations~\cite{or,gefen}. The exact diagonalization method has 
the advantage of taking the full effect of the correlations into 
account, but can treat only small systems. The HF approximations lose 
some of the correlation effects, but are able to handle larger 
systems.

In this paper we would like to answer some of the remaining open 
questions regarding the addition spacing distribution: (i) How does 
it depend on the dimensionless conductance $g$? (ii) Does it depend 
on the charge distribution in the dot (i.e., a uniform distribution 
due to a homogeneous background vs. accumulation of charge at the 
boundaries)? (iii) Do larger system sizes where up to a hundred 
electrons may be added to the dot change qualitatively the results 
obtained in the previous studies where only several electrons were 
added per dot?

In order
to answer these questions we study the Bunimovich stadium which is a 
canonical example of chaotic system~\cite{gutzwiller}. The addition 
spectrum of the stadium is calculated for different values of $g$, in 
the presence of a positive background and its absence, using a 
self-consistent Hartree-Fock approximation. We consider spinless 
electrons, since from experiment~\cite{sba,shw,charlie} and from our 
exact diagonalization calculations~\cite{ber} we can argue that at 
$r_s \sim 1$ the role
of spins in determining the addition spectrum is not an important 
one. 

We find that for any value of $g$ the addition spectrum distribution 
is Gaussian for values
of $r_s \sim 1$. For smaller values of $g$ the distribution becomes 
Gaussian at smaller values of $r_s$.
The dependence of the distribution width on $g$ can be summed up by a 
scaling function given later on. For $g \rightarrow \infty$ the width 
grows only moderately even at $r_s \sim 1$, while for higher values 
of $g$ the distribution is much wider. We find that the background 
plays an important role in determining the distribution of the 
addition spectrum.
The absence of a positive background enhances the width of the 
distribution due to charge accumulation on the boundaries as 
predicted in Ref.~\cite{Blanter}.

To perform our numerical calculations we chose a simple two 
dimensional tight-binding model, namely a square lattice with 
nearest-neighbors interaction. By using a single spatial index to 
label the sites, we can write the Hamiltonian as: \begin{eqnarray} 
\label{eq:hamiltonian}
H&=&H_{0}+H_{int}\\ \nonumber
H_{0}&=&\sum_{j}\epsilon_j a_{j}^+a_{j}
-V \sum_{<j,k>}(a_{j}^{+}a_{k}+a_{k}^{+}a_{j}) \label{h0} \\ 
H_{int}&=&\sum_{j>k} U_{jk} (a_{j}^+a_{j}-K) (a_{k}^+a_{k}-K)\ 
,\label{hint} \end{eqnarray} where $\epsilon_{j}$ is the on-site 
energy, $V$ is the constant hopping matrix element and $\langle 
j,k\rangle$ denotes the sum over the nearest-neighbors. $H_{int}$ 
contains the interaction among the electrons, which we chose as the 
unscreened Coulomb interaction $U_{jk}=U/|r_j-r_k|/b$, with $b$ 
representing the lattice spacing and $U=e^2/b$ the interaction 
strength between electrons located on nearest-neighbor sites. Thus, 
$r_{s}\sim\sqrt{\pi N/n}(U/8V)$, where $N$ is the number of sites and 
$n$ the number of electrons. The constant term $K$ which appears in 
the interaction term represents a constant positive charge 
background. Setting its value equal to zero corresponds to an 
electrically isolated dot, while setting $K=n/N$ assures the global 
charge neutrality corresponding to a dot closely coupled to a gate or 
screened by the environment.

To describe completely the system we need also to specify the 
boundary conditions which define the shape of the dot. We studied a 
Bunimovich quarter stadium, see Fig.~\ref{fig:stadium}, to avoid 
degeneracies due to space symmetries. The classical motion inside 
this region is chaotic and the quantum dynamic shows the typical 
quantum chaos signatures, as the RMT statistics (avoided-level 
crossings and Wigner-Dyson level-spacing distribution) or the 
presence of quantum eigenstates scarred along the unstable classical 
orbits~\cite{scars,bird}. This choice allowed us to study the 
quasi-ballistic regime of high dimensionless conductance but avoiding 
the highly degenerate condition related to the study of regular dots. 

The addition spectrum of the dot has been studied numerically by 
solving the Hartree-Fock problem in various parameter ranges. We 
decoupled the Coulomb interaction in a direct and an exchange term as 
described in~\cite{shf,vojta}. The Hamiltonian in the HF 
approximation reads \begin{eqnarray} \label{eq:shf} 
H_{HF}=H_{0}+\sum_{j\neq k} a_{j}^+a_{j}U_{jk}\langle a_k^+a_k 
\rangle_0
-\sum_{j\neq k} a_j^+a_k U_{jk} \langle a_k^+a_j \rangle_0+const.\ , 
\end{eqnarray}
where $\langle\ldots\rangle_0$ denotes the average on the ground 
state which is calculated self-consistently. 

We studied the addition spectrum by changing the number of electrons 
from $n=20$ to $n=120$ and repeating the calculation for $20$ 
different realizations of the on-site energies. The dimensionless 
conductance of the sample is calculated from the non-interacting 
participation ratio $I= 10^{-2} \Omega^{-1} \sum_{n=20}^{120} 
\sum_{j} |\Psi_n(\vec r_j)|^4$ (where $\Psi_n$ is the n-th 
non-interacting eigenvector and $\Omega$ is the volume) by using the 
relation given in Ref.~\cite{pa} \begin{equation} 
\label{eq:conductance}
g = 3 (\pi(I-3))^{-1} \sum_\mu (\omega_1/\omega_\mu)\ , 
\end{equation} where $\omega_\mu$ are the eigenvalues
of the diffusion equation with the Neuman boundary condition for the 
stadium. It turns out that for a clean stadium or for small disorder 
$g$ acquires negative
values which indicates that, although the energy spectrum follows RMT 
predictions, the eigenvectors are not yet fully random. Thus, in order 
to obtain a really random system we added two types of disorder: (i) 
on-site disorder, where the energy of each site is chosen randomly 
between $-W/2$ and $W/2$ and (ii) strong scatterers, where with a 
certain probability the site energy is set to zero or to a very large 
value (in our case $\epsilon_{j}=100V$). 

In Fig.~\ref{fig:distr} we show the results for the spacing 
distribution for two different values of on-site disorder: $W=3V$ 
corresponding to $g \sim 1$ and $W=V$ corresponding to $g \sim \infty 
$ (the conductance strongly increases when the value of $W$ 
decreases, but $W=V$ lies close to the region where 
Eq.~(\ref{eq:conductance}) fails, thus we could not evaluate the 
exact value of $g$ but only asses, by extrapolation, that it is very 
large).
We studied the case with (c,d) and without (a,b) a compensating 
background, and we built the distributions by averaging over the $20$ 
realizations of disorder and over the number of electrons. 
From this figure we can clearly realize that by 
increasing the interaction strength $U$ we obtain a transition from a 
Wigner-Dyson surmise (solid thin line) to more symmetric and broad 
distributions which are well described by Gaussians (see for example 
the solid thick lines which fit the $U=1.5$ distributions). This 
behaviour is present for both strengths of disorder and for the different 
backgrounds. It is clearly seen that in the absence of a positive 
background the distributions for the same value of $g$ become 
broader, as predicted in Ref.~\cite{Blanter}. This can be understood 
intuitively by noting that the electronic density in the presence of 
background is uniform, while in the absence of background there is a 
charge accumulation on the boundary (we have verified this directly 
from our calculation). Thus, without background the density is 
inhomogeneous and its effective value within the sample is lower, 
leading to a stronger influence of the interaction. The influence of 
$g$ is also clearly demonstrated in the figure. Smaller values of $g$ 
lead to a broader distribution for any kind of background. 

A more quantitative representation can be obtained by calculating the 
distributions momenta. In Fig.~\ref{fig:d2width} we show the 
dependence on the interaction strength of: (a) the average spacing 
$\langle\Delta_{2}\rangle$; (b) the standard deviation 
$\langle\delta\Delta_{2}\rangle
\equiv \sqrt{\langle(\Delta_{2}-\langle\Delta_{2} 
\rangle)^{2}\rangle}$ and (c) the normalized deviation 
$\langle\delta\Delta_{2}\rangle/\langle\Delta_{2}\rangle$. In this 
figure we show the data corresponding to the values of parameter used 
in Fig.~\ref{fig:distr} and also the results obtained by using the 
strong scatterers disorder, where we distributed the scatterers 
randomly in the dot with a $5\%$ probability. This allowed us to 
obtain a high value of the conductance in a case which is close to 
the clean case but where we can successfully use 
Eq.~\ref{eq:conductance}.

Before discussing these results we need to asses their accuracy. The 
HF approximation is less and less accurate as the interaction 
strength increases. In fact we were not able to obtain reliable 
results, and not even convergence of the method, for $U/V$ greater or 
close to $2$, depending on the strength of the disorder. We realized 
that one of the first indicator of the lose of accuracy is the 
appearance of {\em too large} fluctuations in the addition spectrum. 
Thus to estimate the numerical errors affecting the results of 
Fig.~\ref{fig:d2width} we fitted the distributions with a Gaussian 
function and we considered the long-tail deviations from the Gaussian 
as numerical errors. Naturally this procedure, which we used only for 
value of the $U/V$ larger than $0.5$, where the distributions have 
already lost the Wigner-like asymmetry of the non-interacting case, 
can hide some characteristics of the phenomenon. In fact we realized 
that the presence of the tails depends on the kind of disordered we 
used and that, for example, it is more pronounced in the clean dot 
case (not shown here) or in the strong scatterers disorder. We think 
that this effect could be related to the presence of scarred 
eigenstates of the system. These states in fact, for their strongly 
inhomogeneous charge distribution, can produce large charging
fluctuations~\cite{stopa,preparation}. By considering the tails of 
the distribution as exclusively produced by the numerical 
approximation we obtained an overestimate of the error, but in all 
the cases shown in Fig.~\ref{fig:d2width} we obtained error bars 
smaller or comparable with the size of the symbols we used, thus we 
decided not to show the error bars in the figures. Moreover this 
result confirms that at small values of the interaction the spacing 
distribution is already very close to a Gaussian. 

As expected the average spacing (Fig.~\ref{fig:d2width}a) does not 
depend on the value of $g$. With the background 
$\langle\Delta_{2}\rangle$ is linear with $U$, as expected for a 
constant density system~\cite{ba}, while in the absence of the 
background the average spacing is somewhat suppressed because of the 
accumulation of charge on the boundary.

The standard deviation shows an enhancement as function of the 
interaction strength (Fig.~\ref{fig:d2width}b). The enhancement 
(compared to the theoretical RMT value of $\sqrt{4/\pi-1}\sim 
0.52\Delta$, where $\Delta$ is the mean level spacing) strongly 
depends on the background and on $g$. The previously discussed role 
of the positive background in reducing the standard deviation is 
clearly seen. It is also clear that the standard deviation is smaller 
for larger values of $g$.

In Fig.~\ref{fig:d2width}c we see that the normalized deviation for 
the cases in which a positive background exists saturates at values 
of $U\sim 1.5V$. The normalized values saturate at values between 
$10\%$ (for $g\rightarrow\infty$) and $20\%$ (for $g=1$) of the 
average spacing, in agreement with previous results~\cite{sba}. Even 
for $g\rightarrow\infty$
the distribution seems to follow a Gaussian distribution, with an 
interaction independent width already at $U=1.5V$ corresponding to an 
average $\bar r_s = 0.7$ (where $\bar r_s = \sqrt{\pi N} (U/8V) 
\int_{n_1}^{n_2} n^{-1/2}/(n_2-n_1) = \sqrt{\pi N} (U/4V) (\sqrt{n_2} 
- \sqrt{n_1})/(n_2-n_1)$, with $n_1=20$ and $n_2=120$). In the 
absence of background the normalized deviation for the same value of 
$g$ is larger. Moreover, there is a minimum of the normalized 
deviation at $U \sim V$ after which this quantity starts growing. 
This is connected with the fact that at $U \sim V$ the average 
fluctuation $\langle\Delta_{2}\rangle$ starts to deviate from a 
linear dependence on $U$.

For the positive background case we are able to give an empirical 
scaling function for the standard deviation which describes rather 
well the influence of the dimensionless conductance $g$. From 
Fig.~\ref{fig:d2width}b it is clear that the standard deviation 
depends on $g$ and on the interaction strength $U$. In our model 
Hamiltonian we have five independent parameters: $W$, $U$, $V$, $n$ 
and $N$. Out of these parameters one may create a set of four 
independent dimensionless quantities $g$, $U/V$, $r_s$, and 
$\Delta/V$, and the standard deviation can be recast into the 
dimensionless form $\langle \delta \Delta_{2}\rangle^{*} =
\langle\delta\Delta_{2}\rangle/\sqrt{4/\pi-1}\Delta$. We observed 
that, by dividing the dimensionless deviation by $1+U/V\sqrt{g}$ all 
the curves of the same averaged electronic density seem to collapse on 
top of each other (see Fig.~\ref{fig:scaling}). This hints to a 
scaling of the dimensionless deviation $\langle \delta 
\Delta_{2}\rangle^{*} =
(1+U/V\sqrt{g})F(U/V,r_s)$, where $F(U/V,r_s)$ is some function of 
interaction strength and density.
It is important to note that this scaling form does not agree with 
the expectations of perturbation theory. First, for $g \rightarrow 
\infty$ there should be no dependence on $U$ for weak interactions,
which is clearly not the case. Also for a uniform distribution of 
charge one expects the corrections to the standard deviation to 
follow $1/g$~\cite{Blanter} and not $1/\sqrt{g}$. This $1/\sqrt{g}$ 
dependence has been independently seen also in the work of Walker et. 
al.\cite{gefen}.
Of course one may
blame these discrepancies on the HF approximation, but we note that 
for the relative weak interaction strength discussed here one expects 
HF to work rather well. One may also note that we are not able to 
recast the interplay
between interaction strength and density as function of one combined 
parameter $r_s$. This may be the result of the high ratio of single 
electron spacing to the band width in our model. 

In conclusions, the spacing distribution of spinless electrons in 
interacting quantum dots is Gaussian once $r_s$ is of order of one, 
independently of parameters such as dimensionless conductance and the 
positive background charge. Nevertheless, these parameters have a 
strong influence on the onset of the Gaussian distribution and its 
width. This might explain the considerable difference in the width of 
distribution seen between different
experiments~\cite{sba,shw,charlie}, although it is not clear whether 
the physical parameters of the different dots are indeed different. 

Useful discussions on the addition spectrum of quantum dots with O. 
Agam, B. L. Altshuler, A. Auerbach, Y. Gefen, C. M. Marcus, A. D. 
Mirlin, D. Orgad, O. Prus and U. Sivan are gratefully acknowledged. 
L.B. is grateful to G. Grosso and T. Wojta for useful suggestions 
about the Hartree-Fock method. We would like to thank the Israel 
Science Foundations Centers of Excellence Program for financial 
support.
\begin{figure}[tbp]
\caption[]{The
boundary condition used to obtain a chaotic dot without symmetries. 
In our calculations we used the following dimensions: $a=12$, $R=12$, 
$b=1$ which result in $N=265$.} 
\protect\label{fig:stadium} 
\end{figure}
\begin{figure}[tbp]
\centerline{\vbox{
\hbox{
     }
\hbox{
     }
                 }
            }
\caption[]{Distributions of the addition spectrum fluctuations. We 
show $P((\Delta_{2}-e^{2}/C)/\langle\Delta_{2}-e^{2}/C\rangle)$ as 
function of the interaction strength $U$ using lines with different 
patterns. The solid thin line indicates the prediction of the RMT 
theory valid in the non-interacting case and the solid thick line a 
Gaussian fit of the $U=1.5 V$ case. The four graphs correspond to 
$W=1$ (a,c) and $W=3$ (b,d) cases without and with a positive 
background, respectively. The distributions are obtained averaging 
over $20$ realizations of the on-site disorder and over the number of 
electrons ranging from $n=20$ to $n=120$. $V=1$} 	 
\protect\label{fig:distr} 
\end{figure}
\begin{figure}[tbp]
\caption[]{The 
average spacing $\langle\Delta_{2}\rangle$ (a), the standard 
deviation of the spacing $\langle \delta \Delta_{2}\rangle$ (b), and 
the normalized deviation $\langle \delta \Delta_{2}\rangle/ 
\langle\Delta_{2}\rangle$ (c), as function of the interaction 
strength $U$ for different disordered stadia: $W=1$ (circles), $W=3$ 
(boxes), and $5\%$ point scatterers (diamonds) with (dashed lines) 
and without background (solid lines). $V=1$} 
\protect\label{fig:d2width}
\end{figure}
\begin{figure}[tbp]
    \centerline{\hbox{
}
}
\caption[]{The dimensionless deviation $\langle\delta 
\Delta_{2}\rangle^{*}$ (a) and $F(U/V,r_s)$ (b) as function of the 
interaction strength $U/V$ for different disordered stadia with 
background: $W=1$ (circles), $W=2$ (triangles), $W=3$ (boxes), and $5\%$ point 
scatterers (diamonds). The empty symbols correspond to averaging over 
the interval $n=20 \div 120\ (\bar
r_{s}=0.47 V)$ while the filled symbols to $n=20 \div 50\ (\bar 
r_{s}=0.6 V)$. The values of $g$ are shown in the labels.} 
\protect\label{fig:scaling}
\end{figure}

\end{document}